%% file: main.tex
\documentclass[sigconf]{acmart}

\setcopyright{none}
\renewcommand\footnotetextcopyrightpermission[1]{}
\usepackage{subcaption}
\usepackage{makecell}

\AtBeginDocument{%
  \providecommand\BibTeX{{%
    \normalfont B\kern-0.5em{\scshape i\kern-0.25em b}\kern-0.8em\TeX}}}

\begin{document}

\title{Predicting Individual Treatment Effects of Large-scale Team Competitions in a Ride-sharing Economy}


% \author{Authors}
% \affiliation{Institute for Clarity in Documentation}
% \email{aaa@corporation.com}

\author{Teng Ye$^{1,3}$, Wei Ai$^2$, Lingyu Zhang$^3$, Ning Luo$^3$, Lulu Zhang$^3$, Jieping Ye$^{1,3}$, Qiaozhu Mei$^1$}
\affiliation{
  \institution{$^1$ University of Michigan, Ann Arbor,$^2$ University of Maryland, College Park $^3$ AI Labs at Didi Chuxing}
}

\email{tengye@umich.edu, aiwei@umd.edu, {zhanglingyu,luoning_i,zhanglulululu,yejieping}@didiglobal.com, qmei@umich.edu}

\fancyhead{}
% \renewcommand{\shortauthors}{Preprint}
\begin{abstract}
  Millions of drivers worldwide have enjoyed financial benefits and work schedule flexibility through a ride-sharing economy, but meanwhile they have suffered from the lack of a sense of identity and career achievement. Equipped with social identity and contest theories, financially incentivized team competitions have been an effective instrument to increase drivers' productivity, job satisfaction, and retention, and to improve revenue over cost for ride-sharing platforms. While these competitions are overall effective, the decisive factors behind the treatment effects and how they affect the outcomes of individual drivers have been largely mysterious. In this study, we analyze data collected from more than 500 large-scale team competitions organized by a leading ride-sharing platform, building machine learning models to predict individual treatment effects.  Through a careful investigation of features and predictors, we are able to reduce out-sample prediction error by more than 24\%. Through interpreting the best-performing models, we discover many novel and actionable insights regarding how to optimize the design and the execution of team competitions on ride-sharing platforms.  A simulated analysis demonstrates that by simply changing a few contest design options, the average treatment effect of a real competition is expected to increase by as much as 26\%.  Our procedure and findings shed light on how to analyze and optimize large-scale online field experiments in general. 
\end{abstract}

\begin{CCSXML}
<ccs2012>
<concept>
<concept_id>10002944.10011123.10011131</concept_id>
<concept_desc>General and reference~Experimentation</concept_desc>
<concept_significance>500</concept_significance>
</concept>
</ccs2012>
\end{CCSXML}

\ccsdesc[500]{General and reference~Experimentation}

\keywords{Individual treatment effect, team competition, field experiment, sharing economy, machine learning}

\maketitle

% ------------ Sections ------------

\input{sections/1-introduction.tex}
\input{sections/2-related_work.tex}

\input{sections/3-dataset.tex}

\input{sections/4-features.tex}

\input{sections/5-prediction.tex}

\input{sections/6-discussion.tex}

\input{sections/7-implication.tex}

\input{sections/8-conclusion.tex}

\section*{Acknowledgments}
We thank Yan Chen for designing the team competition experiments and Hongtu Zhu for helpful discussions. This work is funded in part by the DiDi Research Partnership with the University of Michigan. Jieping Ye and Lingyu Zhang's work is in part funded by the National Key Research and Development Program of China under grant 2018AAA0101100. Qiaozhu Mei's work is in part supported by the National Science Foundation under grant no. 1633370.

\begin{table*}
\caption{\textbf{Examples of Features with Detailed Description}}
\label{tab:features_detailed}
\begin{tabular}{c}
{\includegraphics[width=0.98\textwidth]{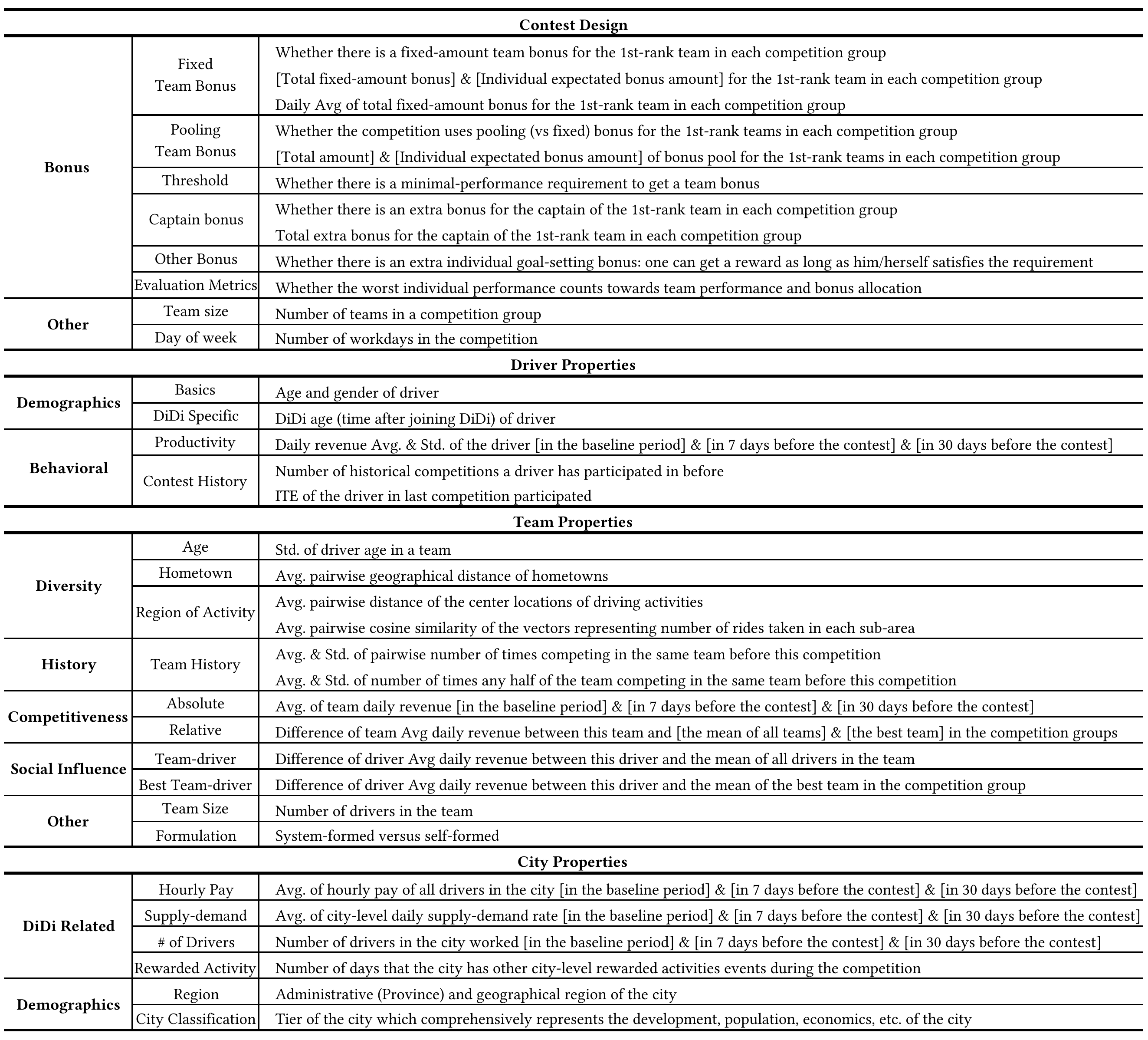}}
\end{tabular}
\end{table*}

\bibliographystyle{ACM-Reference-Format}

\bibliography{team_competition}

\input{sections/supplement.tex}

\end{document}

%% file: sections/1-introduction.tex
\section{Introduction}
The rise of the sharing economy has brought dramatic changes to work and life in modern society. The financial benefits and work schedule flexibility offered by online ride-sharing platforms, such as Uber, Lyft, and Didi Chuxing, have attracted tens of millions of drivers to serve as ride providers. While the drivers enjoy all the values of the ride-sharing economy \cite{chen2019value}, they commonly complain about new barriers to job satisfaction and retention, such as working alone, having few bonds with colleagues, no clear career paths, and a lack of a sense of achievement (e.g., \cite{ny_2017_gig}). How to retain and incentivize service providers to better cover the dynamics of demand has also been a critical problem for the platforms.

Team competitions, practices rooted in social identity theory \cite{akerlof2000economics} and contest theories \cite{vojnovic2015contest}, have been recognized as a potential cure for the pain on both sides. Through competing as teams, drivers are able to (1) build team identity and social bonds with teammates; (2) create a sense of achievement by winning a competition; and (3) increase their satisfaction and performance at work \cite{ai_working}. The increase in driver productivity often outweighs the cost of organizing and providing financial incentives for these competitions, which creates a win-win situation for both the drivers and the platform.  

Indeed, Didi Chuxing (DiDi), one of the world's leading ride-sharing companies, has launched recommender systems to help their drivers form teams and has organized many financially rewarded team competitions to enhance their satisfaction and productivity \cite{zhang2019recommendation}. In 2018 alone, more than 1,400 team competitions were successfully held across 180 cities, which together involved more than 1 million drivers, who provided 130 million rides. These competitions have yielded promising outcomes overall: the average return on investment is larger than 5, indicating that the increased platform revenue through these competitions is five times the cost.

Behind the overall success, however, plenty of unknowns, pitfalls, and challenges remain. There is huge heterogeneity among the cities, the competitions, the teams, and the drivers. Such heterogeneity produces variation in outcomes (or the treatment effects of these experiments):  \textit{What types of drivers and teams} benefit more from team competition? \textit{What competition designs} better increase driver performance? \textit{In what context} is a competition more likely to be effective? Why does a design work \textit{in one city but not in another}? Understanding how these factors predict the outcomes of individual drivers would not only help the platforms find the optimal design of team competitions for different populations of drivers, but would also help them generalize the success to new contexts. 

Addressing these questions is challenging not only for human operational practitioners but also for data mining algorithms. First, it is intrinsically difficult to measure the causal effect of experiments, which requires a careful definition of individual outcome measures and targets of prediction. Second, the variable space to capture driver, team, contest, and context characteristics is high-dimensional, with complex relationships among them. Identifying the potential predictive factors calls for sophistication in both domain knowledge and data analytics.  Third, the large-scale data involve millions of drivers and transactions and many real-world contexts, requiring the prediction algorithms to be scalable and interpretable. 

In this paper, we take a systematic approach to address these challenges. We formulate the problem as a task to predict the treatment effects of a team competition on \textit{individual drivers}, to which we apply both linear and non-linear machine learning models. Combining insights from both business practice and literature on virtual teams and team competition, we construct a large variety of features and train the prediction model using the data of hundreds of large competitions and half a million drivers. The objective of this study is not to prove the causal effect of team competition but to predict individual driver’s performance in out-of-sample/future competitions. The former is analyzed in an earlier study based on a rigorously randomized field experiment (with no self-selection or pre-participation) using formal econometric analysis \cite{ai_working}.

Evaluated on out-sample competitions, the best-performing model is able to reduce the prediction error from the baseline by 24.50\%. A careful interpretation of the models reveals intriguing predictive power of many factors (for individual treatment effects): some are intuitive, such as team homophily, social influence, supply-demand ratio, and weather conditions; some are rather surprising, such as team diversity, pre-contest activities, and the design of monetary incentives; and many of them have never been reported in the literature.  Some of the factors are directly actionable in business practice, and a simulation analysis demonstrates that by simply varying several contest design options, one is expected to increase the average treatment effect of a competition by as much as 26\%.  

To summarize, we make the following major contributions:

\begin{itemize}
\item We present the first study of individual treatment effects of team contests in a sharing economy.  While existing work measures the average effect of an experiment, we analyze heterogeneous, per-driver outcomes across many experiments. 

\item We define a robust estimation of individual treatment effects and formulate a novel approach to predicting individual treatment effects through machine learning. 

\item We train effective machine learning models on large-scale data collected from hundreds of historical experiments, which combine a comprehensive set of features of individual drivers, teams, contest designs, and experimental environments, and we evaluate the models on out-sample experiments. 

\item We reveal the predictive power of a variety of factors for the outcome of individual drivers, most of which are novel. 

\item We identify actionable implications for business practice and demonstrate significant potential improvements in experimental outcomes by varying several contest design options. 

\end{itemize}

%% file: sections/2-related_work.tex
\section{Related Work}
This study is related to the following lines of literature: 

% \paragraph{Sharing economy}
\textbf{Sharing economy.} A growing literature investigates the socio-economic effects on and consequences of ride-sharing platforms, such as Uber and Lyft \cite{zhang2017quasi}. Inspired by the findings in ~\cite{hamari2016sharing} that economic gains positively influence people's intention to participate, a stream of work quantifies the positive effect of financial incentives, such as subsidy~\cite{fang2019prices}, on improving supply-demand efficiency. Our study adds to this literature by investigating the effect of rewarded team competitions on service provision in a ride-sharing economy. 

% \paragraph{Team competition}
\textbf{Team competition.}
Team competitions have been increasingly applied in online communities, such as crowdsourcing \cite{rokicki2015groupsourcing}, education \cite{scales2016randomized}, online games \cite{ cheng2019makes}, and charitable giving \cite{chen2017does}. It has been shown that team competitions are effective in improving key metrics, such as participation \cite{scales2016randomized}. Data-mining researchers have developed team matching algorithms to ensure team formation of high efficiency, effectiveness, and fairness, taking into account factors such as demographics, social networks, and tasks (e.g., \cite{ai2016recommending,zhang2019recommendation,anagnostopoulos2012online}).

Most of these studies demonstrate the effect of team competitions through either field experiments or analyzing observational data. The former usually estimate the treatment effect at the experiment level, averaged over all treated teams and participants (e.g., \cite{chen2017does,scales2016randomized,rokicki2015groupsourcing}). Studies of the latter have examined team-level properties and their influences on team performance in online games, such as the positive factor of diverse team composition
\cite{cheng2019makes}. To the best of our knowledge, few have aimed to analyze and predict the heterogeneous effect of team competitions on individual team members, especially in the context of the sharing economy.

% \paragraph{Individual treatment effects \& counterfactual analysis} 
\textbf{Individual treatment effects \& counterfactual analysis.} 
Recent work in causal inference and machine learning has focused on a finer granularity -- individual treatment effect (ITE) estimation, citing its potential in precision medicine~\cite{fang2019applying} and online platforms~\cite{makar2019distillation}. Estimating ITE has been done with random forests~\cite{athey2016recursive} and deep neural networks~\cite{shalit2017estimating}, and it has taken into account hidden confounders from network information~\cite{guo2020learning}. We base our analysis on a collection of online controlled experiments \cite{kohavi2013online}. We are able to estimate ITE with difference-in-differences (DID), as the team contests already include randomly selected control groups. We thus focus on the prediction of ITE. 

Another related stream of literature is counterfactual learning, where the focus is to learn what policies maximize some rewards, such as engagement or conversion in online advertising~\cite{bottou2013counterfactual, swaminathan2015selfnormalized}. The counterfactual estimators are typically based on importance sampling. Our paper also examines how policy (which is the contest design in our setting) predicts ITE, but we study the predictors of ITE in a much more complex socio-economic setting.

%% file: sections/3-dataset.tex
\section{Problem Setup}
\label{sec:dataset}

\subsection{Team Competitions on DiDi}
\label{sec:didi-teams}
\begin{figure}
\centering
\includegraphics[width=0.8\columnwidth]{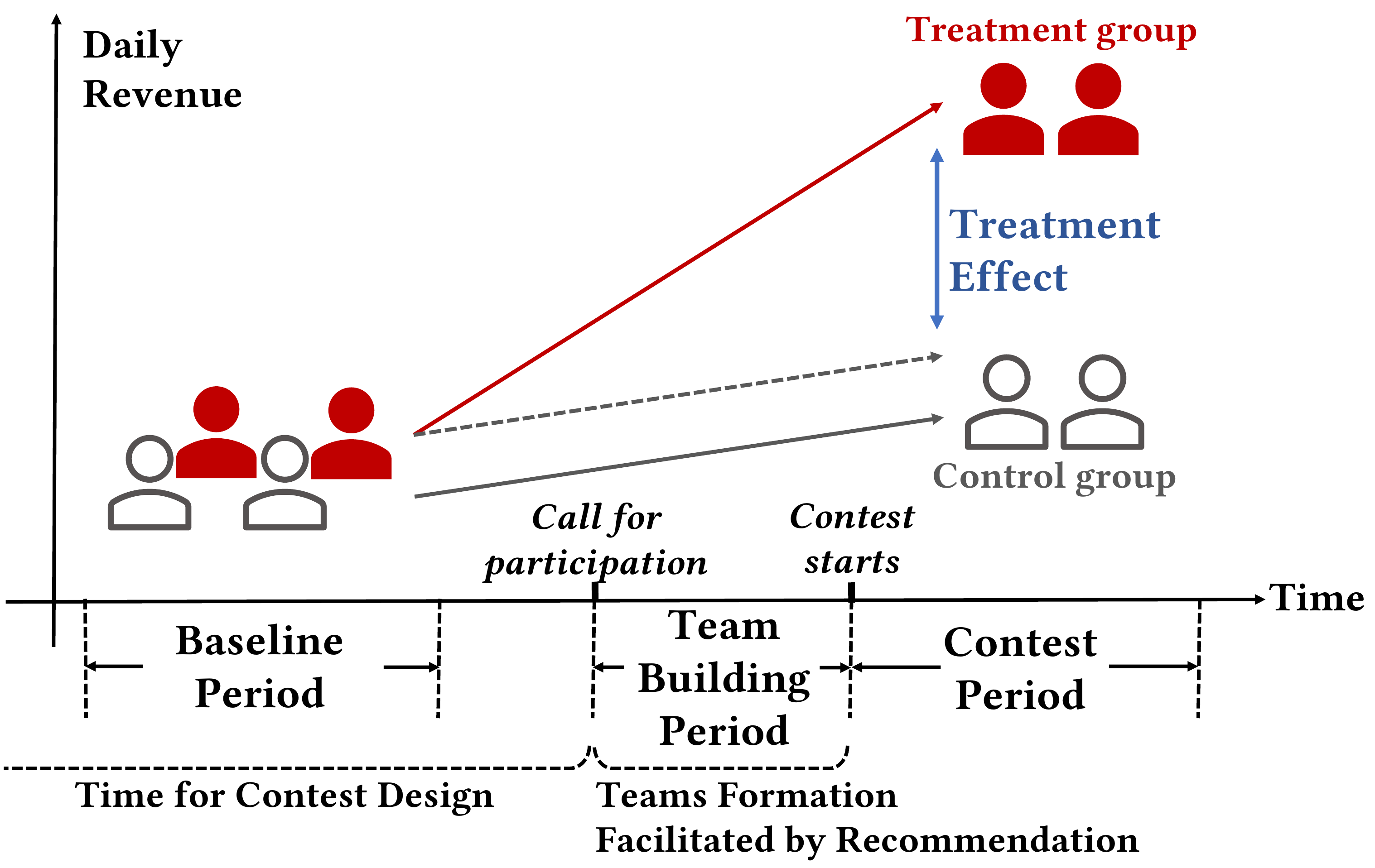}
\vskip -5pt
\caption{Workflow and Treatment Effect of a Team Contest}
\vskip -10pt
\label{fig:periods}
\end{figure}

Since 2017, team competitions (also referred to as team contests) have been widely introduced as driver incentive campaigns in DiDi \cite{ai_working}. A typical team contest is held in one city and consists of two periods: a \textit{team building period} and a \textit{contest period} (see Figure~\ref{fig:periods}).

\paragraph{Team building period}
The team building period starts with a call for participation and usually lasts 3-7 days. During this period, interested drivers sign up for the contest and start teaming up. Drivers can create a new team as captain or join an existing team by invitation, and they can invite other drivers into a team either manually or assisted by a recommender system~\cite{zhang2019recommendation}. All participating teams in one contest have the same size: one captain and 2 to 7 other regular members.

About half of the teams achieve the desired size during the team building period; these are referred to as \textit{self-formed} teams. At the end of the team building period, the system \textit{randomly} selects 90\% of the unteamed drivers and groups them into full-sized teams, which we refer to as \textit{system-formed} teams. The other 10\% are not assigned to any team and will not participate in the competition; they are referred to as \textit{solo drivers}. These solo drivers have the same motivation level (to participate in the contest), productivity, and other demographic properties compared to the drivers who are assigned to teams, and they form a nice group for comparison. The system keeps track of the solo drivers for control.

\paragraph{Contest period}
Both self-formed and system-formed teams will compete during the contest period. The teams are further partitioned into smaller contest groups. Each contest group contains the same number of (usually 5) teams of comparable competitiveness, measured by their productivity prior to the contest. A team only competes with other teams within the contest group and will win a cash reward according to its standing in that group. The performance of a team is calculated by summing the productivity of team members, measured by their daily revenue, number of rides, or a combination of both. During the contest period, a driver can check the performance of their team members and competitor teams through a real-time leaderboard. Under these general constraints, every city can choose among finer-grained design options (such as incentive structures). We will summarize these contest design options in Section \ref{sec:contest_design}. 

These team contests have been quite successful in general. During a contest, a driver's daily revenue on average increases by 22\%, and the revenue over investment (ROI, which measures revenue of the platform over cost) is over 5. While the average treatment effect provides an overall picture of the effectiveness of team contests, it is critical to understand the treatment effect on individual drivers to untangle the complex interplays among participants, teams, contest design, and experimental environments. Only through this can the platform optimize their recommender systems and contest designs, provide targeted interventions for different population of drivers, and to generalize the success to new contests, cities, and countries. 

\subsection{Estimating the Individual Treatment Effect}
\label{sec:ite}

We need to first estimate the individual treatment effects before analyzing and predicting them. Estimating the individual treatment effect by itself can be challenging in natural experiments and observational data \cite{athey2016recursive,makar2019distillation,shalit2017estimating}. In our scenario it is easier, as all the competitions followed a rigorous experimental design. 

The individual treatment effect (ITE) refers to the effect of a single team contest on the revenue of an individual driver. In other words, the effect measures how much additional revenue a driver generates by participating in a team competition as opposed to otherwise. Given the competition setting, we estimate the individual treatment effect using a standard difference-in-differences (DID) approach~\cite{angrist2008mostly} in causal inference. The intuition of DID is to first compute the difference in revenue before and during the contest for each driver, aggregate such within-driver differences by treatment status (treatment vs. control), and compare the differences between the two conditions. In our case, the control group is clear - the solo drivers (drivers who are not teamed). We have two possible definitions of the treatment group: (1) drivers in both system-formed and self-formed teams; (2) drivers in system-formed teams only. Ideally, drivers in system-formed teams are the most comparable to solo drivers, as self-formed teams might differ in motivation or pre-contest history, which introduces a potential selection bias. In business operation, however, we do care about making predictions for all drivers. We therefore separately analyze the two scenarios: using ``all teams'' and using ``system-formed teams'' as treatment group. If the results are consistent, that means the estimation of ITE can generalize from system-formed teams to all teams.

Formally, we define $R_{j,T}$ as the average daily revenue generated by driver $j$ in the time period $T$. $T=T_1$ indicates the contest period while $T=T_0$ indicates a \textit{baseline period} before contest starts. $T_0$ is selected as the most recent days prior to the call for participation, conditioned on matching the length and the day(s)-of-the-week of $T_1$. The choice of $T_0$ rules out day-of-the-week confounds on revenue (see Figure~\ref{fig:periods} for illustration). 

The within-driver difference in revenue between the contest period and the baseline period can thus be calculated as
\begin{equation}
\label{eqn:first_diff}
\small
\Delta R_j = R_{j, T_1} - R_{j, T_0}.
\end{equation}

We then aggregate the revenue change in the control group as
\begin{equation}
\label{eqn:common_trend}
\small
\Delta R_\text{control} = \frac{1}{|\text{control}|}\sum_{i\in\text{control}}\Delta R_i.
\end{equation}

Finally, we can obtain the individual treatment effect as
\begin{equation}
\label{eqn:ite}
\small
\Delta R_{j}^{\text{ITE}} = \Delta R_{j} - \Delta R_\text{control},
\end{equation}
for every driver $j$ in a team. If we calculate the average value of the ITE of a given contest, we will get the \textit{average treatment effect (ATE)} of that contest. More precisely, since we can only obtain the ITE of treated drivers (participating in the team contest), the aggregated ITE represents the \textit{average treatment effect on the treated (ATET)}.

\subsection{Predicting the Individual Treatment Effect}

We collect a dataset from all competitions held between January 1, 2018 and August 23, 2018. Contests that did not hold out the 10\% solo drivers are excluded, as we lack the control condition to calculate ITE. We also exclude the contests conducted during the lunar new year, as the supply and demand pattern in that period is irregular. For all selected contests, we collect the demographics and historical activities of all drivers who sign up for the contests, regardless of whether they are in the treatment or control group. Table \ref{tab:datasets} presents the summary statistics of the contests included.

Based on this dataset, given every contest $C_k$, we are able to represent it with a list of \textit{driver-independent} features (such as information about the city and the contest design), $\mathcal{F}_{C_k}$. For every treated driver $j$ in $C_k$, we are able to estimate the treatment effect of $C_k$ on $j$, $\Delta R_{C_k, j}^{\text{ITE}}$. Let the start time of the team contest period of $C_k$ be $t_k$; we represent a driver $j$ with a set of features about their demographics or activities that are observed \textit{before} $t_k$, denoted as $\mathcal{F}_{j, t_k}$. We are also able to represent the team that $j$ joins, $\text{team}(j)$, with a set of features $\mathcal{F}_{\text{team}(j)}$. Note that $\mathcal{F}_{\text{team}(j)}$ could contain aggregated features of its members, or $\mathcal{F}_{\text{team}(j)} \sim g(\mathcal{F}_{i, t_k}|i \in \text{team}(j))$. 

\begin{table}[t]
\caption{Summary of Statistics}
\vskip-10pt
\label{tab:datasets}
\small
\begin{tabular}{l|c||l|c}
\hline
Item & Number & Item & Number\\
\hline 
\# of Cities & 143 & \# of Unique Drivers & 520,611 \\
\# of Contests & 520 &
\# of Cumulative Participation & 887,842\\
\hline                                        
\end{tabular}
\vskip-10pt
\end{table}

Given these notations, we define the problem of predicting the individual treatment effect as finding a function $f(\cdot)$ that maps the feature representations of the contest $C_k$, a driver $j$, and the team $\text{team}(j)$ to the treatment effect of $C_k$ on $j$, that is, 
\begin{equation}
\small
    \Delta R_{C_k, j}^{\text{ITE}} = f(\mathcal{F}_{C_k}, \mathcal{F}_{j, t_k}, \mathcal{F}_{\text{team}(j)}). 
\end{equation}

The prediction problem as defined is intrinsically challenging. First, predicting human behavior is hard given the great complexity in cognition and decision making \cite{subrahmanian2017predicting}. Second, $\Delta R_{C_k, j}^{\text{ITE}}$ as defined is essentially a ``change'' in behavior, which is harder to predict than the behavior itself. Moreover, the huge heterogeneity among drivers, teams, contests, time, and environments results in a wide variation in the ITE. These challenges call for a careful selection of features and predictors. In the following sections, we show how to extract the feature representations of $\mathcal{F}_{C_k}$, $\mathcal{F}_{j, t_k}$, and $\mathcal{F}_{\text{team}(j)}$, and how to find the function $f(\cdot)$ through a machine learning approach.

%% file: sections/4-features.tex
\section{Predictive Features}
\label{sec:feature-exploration}

Our comprehensive dataset presents unprecedented opportunities to measure a wide portfolio of conditions related to the driver, the team, the contest, and the experimental environment. In this section, we characterize these conditions as informative features, generated based on the theoretical insights from the literature on contest theory, social identity theory, and virtual teams, as well as the domain knowledge from the operational practitioners at DiDi.

\subsection{Contest Design}
\label{sec:contest_design}
We start with contest design features, such as the winning condition and the prize structure. This set of features determine the utility function of the participants and directly affect their motivation and efforts devoted. Currently, the platform relies on their intuitions to decide contest designs. They are eager for actionable insights and guidance on how to optimize these designs. Apart from execution options such as team size, contest-group size, and timing, we build upon the theoretical inferences in contest theory or social identity theory to describe the incentive mechanisms in contest design.

For example, how to allocate the prizes in a contest group? Give them all to the best-performing team or split over several placements? Although this question has been analyzed in contest theory: under certain assumptions, rewarding the best in the contest group is the optimal strategy \cite{moldovanu2007contests}, it is seldom tested in field. We code the team bonus for each of the top 5 teams in a contest group.

\subsection{Driver Properties}
This set of features capture the demographics and behavioral patterns of a driver before the contest, which we assume would affect the outcomes. To depict driver behavioral patterns before competitions, we retrieve drivers' daily revenue, daily number of rides, and daily hours on the platform, each in three periods: the \textit{baseline} period (see Section \ref{sec:ite} and Figure \ref{fig:periods}), 7 days before the contest starts, and 30 days before the contest starts. These features are designed to capture the most comparable activities to the contest period, the most recent activities before contest, and the longer-term work habits. We also collect driver demographics, such as age, gender, and number of months on platform (i.e., DiDi age).

\subsection{Team Properties}
This set of features are related to team-level characteristics that may significantly influence the behaviors of a member. Apart from basic team characteristics (e.g., size), we investigate \textit{team diversity}, \textit{team history}, \textit{team competitiveness}, and \textit{the influence of team on individual driver}, drawing upon previous literature  \cite{wegge2008age,ai2016recommending,pinsonneault1999electronic,muchnik2013social}.

\begin{figure}[htbp]
\centering
\vskip -10pt
\includegraphics[width=0.85\columnwidth]{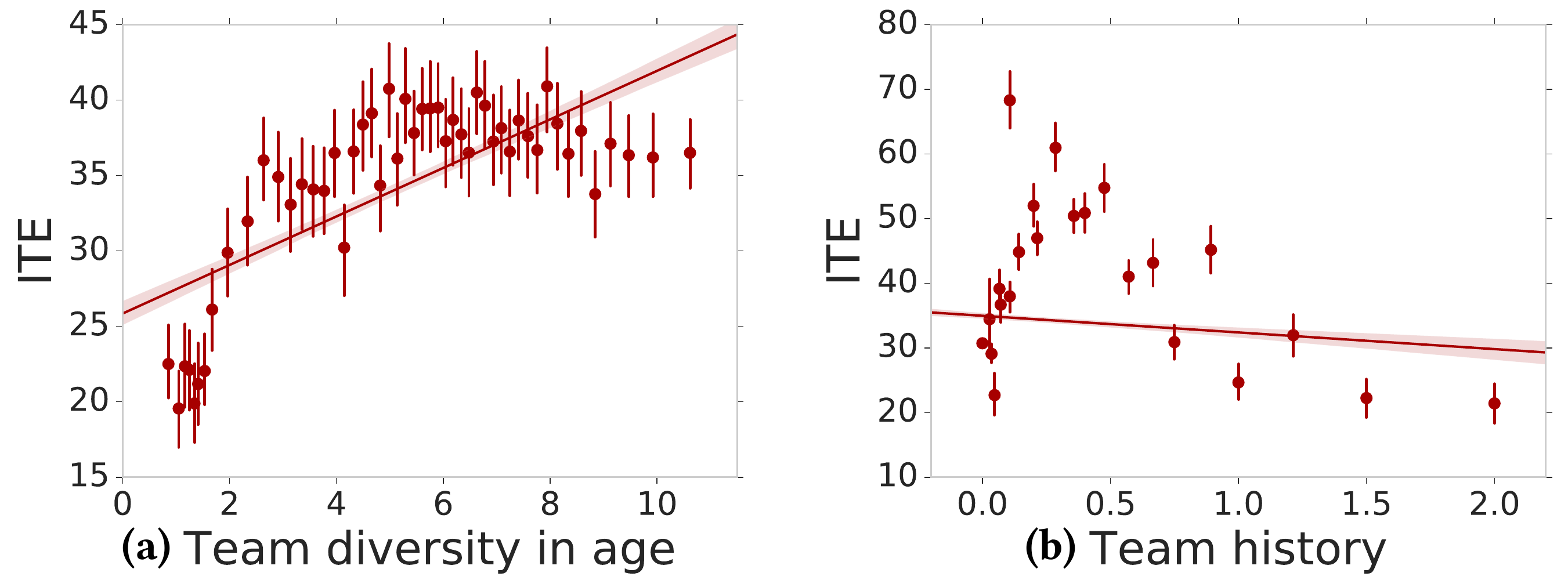}
\Description{}
\vskip -10pt
\caption{Relationship between Features and ITE}
\vskip -10pt
\label{fig:sec4_features}
\end{figure}

For example, we capture team diversity from three aspects: \textit{age diversity}, \textit{hometown diversity}, and \textit{diversity in activity region}. As illustrated by Figure \ref{fig:sec4_features}a, age diversity is shown to be a potential strong predictor of ITE. For another example, to depict \textit{team history}, we calculate the average number of times that any two teammates have been in the same team before \textit{this} competition. While literature has reported both the positive and the negative effect of team history on team performance~\cite{pinsonneault1999electronic}, Figure \ref{fig:sec4_features}b shows that the relationship between team history and ITE follows an inverse-U shape: no history and too much history could be equally harmful! Teams perform the best when on average half of the pairs of drivers have been teammates before, or translated to roughly 70\% old members and 30\% new members if a team is built on a previous team.

\vskip -10pt
\subsection{City Properties}

We also consider the environments where a contest is held, which may influence the motivation and outcome of the contest. 
We describe the status quo of DiDi in the contest city with the number of historical team contests, the number of DiDi drivers, and their average hourly pay. Moreover, we consider general demographics of the city, such as its development level and the province it belongs to. We also retrieve the weather reports of the city during a contest.

A more comprehensive list of features can be found in Table~\ref{tab:features}. Preliminary analysis has identified many correlations between these features and the ITE, although we only show two of them due to the space limit, promising the feasibility of predicting the ITE.

\begin{table}[t!]
\caption{\textbf{Feature Examples} \small{(More Details in Supplement Table \ref{tab:features_detailed})}}
\vskip -10pt
\label{tab:features}
\begin{tabular}{c}
{\includegraphics[width=\columnwidth]{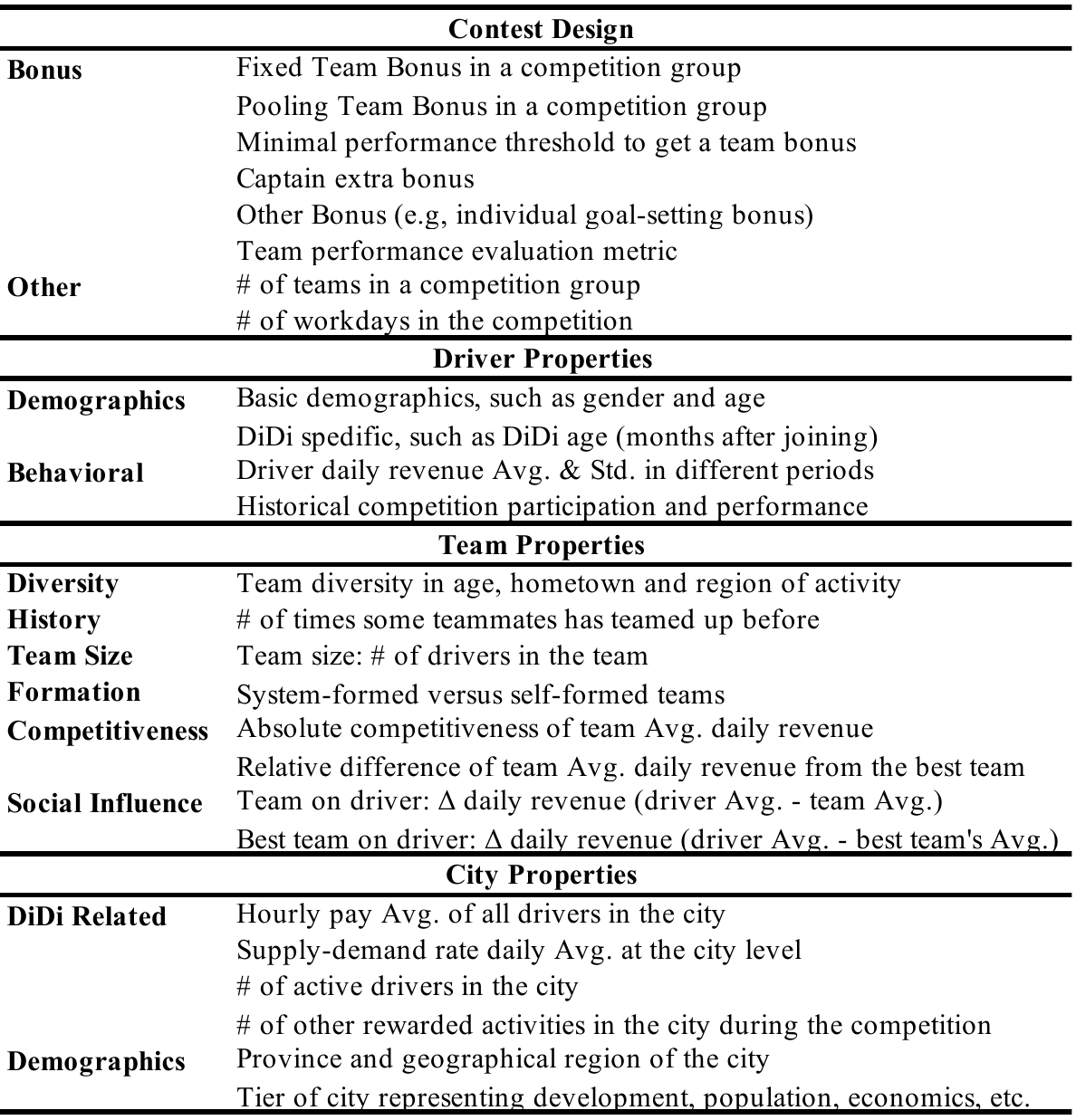}}
\end{tabular}
\vskip -10pt
\end{table}

%% file: sections/5-prediction.tex
\section{Predicting ITE}
\label{sec:prediction}

To what extent can the combination of the factors in Section~\ref{sec:feature-exploration} jointly predict the ITE? Practically, it is also valuable to know how these predictions generalize to out-sample, new competitions. Building machine learning predictors is a desirable solution for both aspects.

\subsection{Model Training and Evaluation}
\label{sec:prediction-problem}

We expand the feature exploration and craft 555 features to represent factors of contests, drivers, teams, and cities (see Table~\ref{tab:features}).
% We expand the feature exploration and craft 555 features based on \textcolor{black}{contest design, driver properties, team properties, and city properties (see example features in Table~\ref{tab:features})}.

We follow the standard practice and split the contests into training, validation, and test sets based on the timing of the contests. Details can be found in the supplement. The performance of a machine learning predictor can be measured with \textit{RMSE}:
\begin{equation}
\small
    \text{RMSE} = \sqrt{\sum_{k,j} \left(\Delta R_{C_k,j}^\text{ITE} - \Delta \hat{R}^\text{ITE}_{C_k,j}\right)^2/
    \sum_{k}N(C_k)},
\end{equation}
where $N(C_k)$ is the number of drivers participating in contest $C_k$. 

There are many machine learning models that can be used for building the predictors. Our main goal is not to optimize the prediction accuracy but rather to understand the effect of individual predictive factors on the target -- the ITE.  Therefore, we consider two objectives in selecting the machine learning algorithms: (1) they should be able to capture the linear and non-linear effects of features and their interactions; (2) they should provide an easy mechanism to interpret the predictive power of individual features. \textcolor{black}{We select two standard and commonly used algorithms. One is Lasso \cite{tibshirani1996regression}. As a linear model, the learned coefficients provide a natural interpretation of the predictive power of features. The other is Gradient Boosted Regression Tree (GBRT)~\cite{friedman2002stochastic}, which can capture the non-linearity and interactions of the features. The feature importances reported by GBRT can help interpret the contributions of different features in predicting ITE.\footnote{We use glmnet 3.0-2 package (https://cran.r-project.org/web/packages/glmnet/index.html) for Lasso, Ridge; scikit-learn 0.20.0 package (https://scikit-learn.org/stable/) for GBRT.} We also train Ridge models \cite{hoerl1970ridge} to verify the robustness of linear models to different regularization. } We did not choose neural networks as it is harder to interpret the importance of individual features with a deep neural network.

\subsection{The Prediction Performance}
We tune the hyperparameters of the machine learning models rigorously based on validation RMSE and report the performance of the models on test set (contests starting in August) in Table~\ref{tab:rmse}. We construct two baseline predictors for comparison. The uniform baseline predicts all ITE as the mean ITE in the training set, while the random baseline draws from a Gaussian distribution estimated from the ITEs in the training set. We separately train the models in two settings, one with drivers in all teams and one with system-formed teams only. From Table~\ref{tab:rmse}, GBRT, Lasso, and Ridge all achieve similar performance, reducing RMSE from the better baseline (Uniform) by up to 24.50\% ($p<.01$) on all teamed drivers and 24.77\% ($p<.01$) on drivers in the system-formed teams only. The consistency between the two settings suggests that the estimation of ITE can generalize from the system-formed teams to all teams.

\begin{figure*}[t]
\centering
\includegraphics[width=0.92\textwidth]{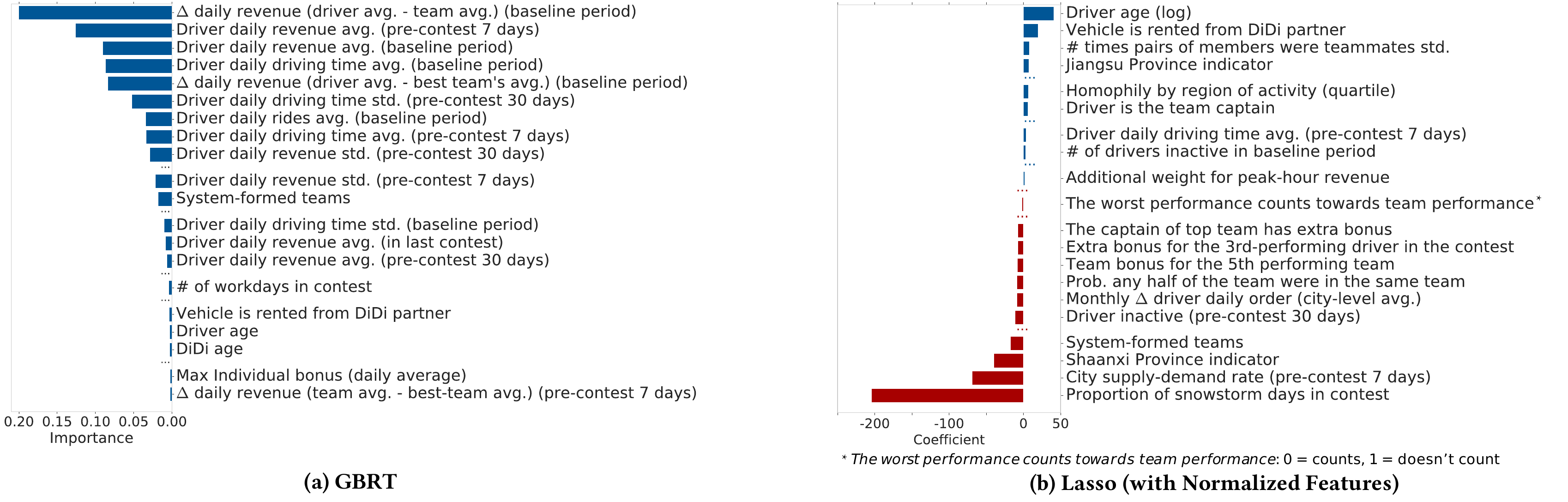}
\vskip-5pt
\caption{Importance Scores of Selected Features from the Best-performing GBRT and Lasso Model for All Teamed Drivers}
\label{fig:gbr_lasso_feature_imp}
\vskip-5pt
\end{figure*}

Note that both GBRT and Lasso are "selecting" features during the training process. By examining the non-zero coefficients in Lasso and the positive feature importances in the GBRT, we can know which salient factors the two models rely on to make predictions. As we can see from Table~\ref{tab:rmse}, the numbers of features selected by the different models are quite different (251 vs. 119). In other words, the two models achieve similar performance based on different sets of features, due to the different model structures.

\renewcommand{\arraystretch}{0.92}
\begin{table}[htbp]%htbp
\vskip -8pt
\caption{Model Performance, Evaluated by RMSE
}
\vskip -10pt
\label{tab:rmse}
\small
\centering
\begin{tabular}{l|ccc|ccc}

\hline 
&\multicolumn{3}{c|}{All-teams Drivers} &\multicolumn{3}{c}{System-formed-teams Drivers}\\
\cline{2-4}\cline{5-7}
Model & Val. R. & Test R. & \# Ftr.& \quad Val. R.~~~~ & ~~Test R. & \# Ftr.\\
\hline
GBRT &  139.19 & 147.96 & 251 & \quad 125.00& 139.67&248 \\
Lasso & 141.75& 148.46 & 119 & \quad 137.25 &141.40 &116  \\
Ridge &142.16 & 150.55&555 & \quad 136.26 & 143.65 & 552 \\
Uniform & - & 195.97 & - & \quad - & 185.66 & - \\
Random & - & 266.34 & - & \quad - & 250.63 & -\\
\hline 
\end{tabular}
\vskip -12pt
\end{table}
\renewcommand{\arraystretch}{1}

%% file: sections/6-discussion.tex
\section{Analyzing Prediction Results}

\subsection{Which Features Predict Treatment Effects?}
We examine the most predictive features nominated from both models. Figure \ref{fig:gbr_lasso_feature_imp}a and \ref{fig:gbr_lasso_feature_imp}b each show 20 selected features from the best-performing GBRT and Lasso models with all-teams dataset. Both all-teams and system-formed-teams datasets produce similar results, and we choose to report the former since we do care about making predictions for everyone when deployed in the operations.

\subsubsection{Contest Environment} \ \\
We start with a set of factors about the environment of the contest. 

\textsc{Weather.} The largest (negative) factor by Lasso for the individual treatment effect is the proportion of snowstorm days during a competition ($p< .01$). This is easy to understand as severe weather conditions would limit travel activities and driving efficiency. 

\textsc{Location. } We observe clear heterogeneity of ITE in different locations. Contests held in certain provinces or cities have significantly higher/lower effects. Basic demographics of the city (such as population) do not appear to be predictive.  The geographical heterogeneity may attribute to other properties of the locations. 

\textsc{Supply-Demand Rate. } Surprisingly, the second largest negative factor identified by Lasso is the supply-demand ratio of the city where a competition is held. Team competitions are more effective in cities of greater supply shortage ($p< .01$). This makes sense, as when supply can't meet demand, more effort of a driver ensures more profit. When supply exceeds demand, even if drivers intend to work harder, they are unlikely to receive more orders. This finding is directly actionable: sharing economy platforms should prioritize incentive-based experiments in areas of a greater supply shortage.

\subsubsection{Driver Demographics} \ \\
\textsc{Young means high? No!} The sharing economy has been commonly perceived as a "young people's business." \footnote{\url{https://www.forbes.com/sites/homaycotte/2015/05/05/millennials-are-driving-the-sharing-economy-and-so-is-big-data/}, retrieved in October, 2019. } However, we find that middle-aged drivers and those who have joined the platform earlier experience greater treatment effects. In both GBRT and Lasso, age of driver is one of the most predictive features. Indeed, in Figure \ref{fig:feature_interpret}a, we observe that the treatment effect of team competitions increases with age, tops among drivers in their 40's, and decreases when drivers are over 50. One possible interpretation may be the high economic pressure on the middle-aged group. ITE also increases with a driver's age on platform. A longer lifespan on the platform indicates more experience and a greater motivation to stay in the business. From Figure \ref{fig:feature_interpret}b, veterans (on DiDi for over a year) have higher ITE ($p<.05$), and the trend does not drop down.  

\textsc{Rental Cars.} Results show that drivers are more productive in competitions when they don't own their vehicles but have rented from a DiDi partner ($p< .01$). One possible reason is that these drivers are more motivated to earn extra rewards to cover the rental cost,  or simply the rental vehicles are in better conditions. 

\subsubsection{Pre-contest Activities} \ \\
The pre-contest activities of a driver show strong predictive power.

\begin{figure}[t]
\centering

\includegraphics[width=0.82\columnwidth]{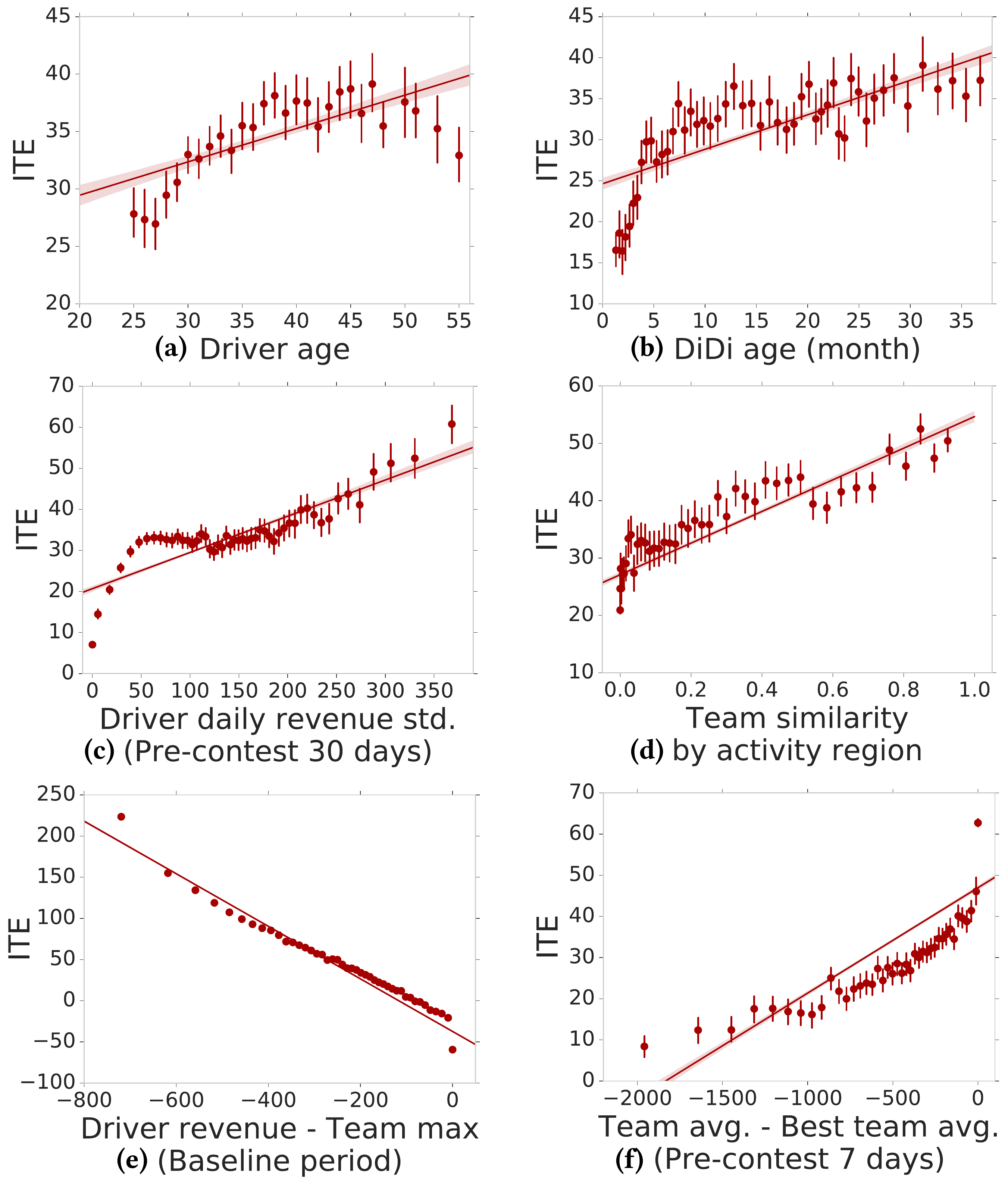}
\vskip -10pt
\caption{Relationships between Features and ITE}
\label{fig:feature_interpret}
\vskip -10pt
\end{figure}

\textsc{Productivity in Previous Competition.} Results (see Figure \ref{fig:gbr_lasso_feature_imp}a) suggest that the individual treatment effect of \textit{this} contest depends on the revenue the same driver received in the previous contest they participated in ($p< .01$). Not surprisingly, if a treatment was effective on someone, the same thing would likely work again. 

\textsc{Productivity Variation.} One of the most surprising factors is the variance in pre-contest activity levels.  Results show that the \textit{standard deviation} of a driver's daily revenue in the 30-day period before the contest has a positive correlation to the treatment effect ($p< .01$) from Figure \ref{fig:feature_interpret}c. Similar effects are observed when productivity is measured by work time or number of orders. When a driver's work habits are irregular, inner-team coordination may drag their behavior towards the social norm. Drivers of a high variation are also likely to be working part-time, and they have more room to improve through the competition. 

\subsubsection{Team Properties} \ \\
Team structures and interrelations between members are also predictive. In social network and organization literature, there are theoretical and empirical discussions about how structural properties affect the functionality of a team or community (e.g., \cite{cheng2019makes,borgatti2009network}). Our analysis provides empirical evidence (in the context of the sharing economy) to these theories while also reveals novel findings. 

\textsc{Homophily.} From Figure \ref{fig:gbr_lasso_feature_imp}b, we observe that \textit{homophily (similarity of team members) by region of activities} is a strong predictor of the treatment effect. This effect is positive and almost linear (see Figure \ref{fig:feature_interpret}d). Previous literature suggests that physical distance (the inverse of homophily) negatively influences the performance of virtual work teams as it reduces shared contextual knowledge, emotional attachment, and non-verbal cues in collaborations among team members \cite{kraut2002understanding}. Our result extends previous work by finding that physical distance is harmful ($p< .01$) even when it requires little coordination and communication to complete the team tasks.

\textsc{System-formed Teams.} The method of team formation is an important predictor in both models. Teams filled by the system on average yield a smaller treatment effect than teams fully formed by drivers ($p< .01$). We note an apparent confound that drivers who form teams without the help of the platform already knew each other: they may be acquaintances in real life (related to homophily) or they may have been teammates in previous competitions.

\textsc{Role of captain.} We find that \textit{team captains} generally have higher ITE than other team members during competitions ($p< .01$, see Figure \ref{fig:gbr_lasso_feature_imp}b). This is intuitive, as drivers who volunteer to be captains are likely to be more dedicated. Another possible explanation is that the captains are ``leading by example''~\cite{hermalin1997economic}.

\textsc{Social influence.} A rather intriguing finding by the GBRT is that social influence, rather than individual behaviors, is a strong predictor of ITE. As shown by Figure \ref{fig:gbr_lasso_feature_imp}a, the top feature measures the difference between the pre-contest productivity of a driver and the average pre-contest productivity of the team. The lower a driver's pre-contest productivity is than the team average, the higher their productivity increases through the team competition ($p< .01$). This desirable outcome may be attributed to how a team functions, as social influence drags the inactive or inexperienced drivers towards the norm \cite{cialdini1998social}. Note that for drivers who were already significantly more productive than their team average, the team may have also dragged them backwards towards the norm. Do these drivers constitute a large proportion in each team? By calculating the difference between the pre-contest productivity of individual drivers and the most productive team member instead of the team average (Figure \ref{fig:feature_interpret}e), we see that most drivers receive a positive social influence, unless they are (or are close to) the most productive ones in their teams (with this difference close to zero).

In contrast, drivers are more motivated when the pre-contest productivity of their team is closer to that of their competitors. As shown in Figure \ref{fig:feature_interpret}f, ITE is higher when the pre-contest productivity of a team is closer to that of the winning team in its contest group. 

These findings provide novel insights for both team formation and contest design: it is desirable to mix drivers with different activity levels, so that the more productive/experienced drivers may help the others and improve team performance. However, such a service role may hinder the motivation of the top drivers and slow down their own productivity, so it is important to provide additional incentives to the helpers. It is also important to match the competitors so that all the teams are competitive in the group.

\subsubsection{Contest Design} \ \\
\label{sec:contest_interpret}
\textsc{More is Less!} Contrary to common sense, our results show that providing more bonuses does not necessarily lead to a better outcome. Specifically, the Lasso model suggests that while in general drivers work harder for high financial rewards, an ill-designed extra bonus could inhibit the treatment effect. For example, when the 5th-performing team (the bottom team in most contests) in a contest group is rewarded, drivers become less motivated as everyone is guaranteed some reward ($p< .01$). In addition, if team captains receive an extra bonus, drivers in general become less productive ($p< .01$). The inequality between captains and members might have shifted the team goal from fighting for team identity to fighting for the captain, reducing the motivation of others. 

\textsc{Inner-team Competition. } Adding enforced within-team competition might hurt the treatment effect: drivers are less productive if the revenue of the worst-performing driver is excluded from calculating team performance and bonus allocation ($p< .01$). Note that without this arbitrary mechanism, there is also implicit, natural competition among team members, as in most contests, the rewards are allocated based on the contributions of members.   

In general, the above findings are directly actionable by contest organizers, to improve the outcomes of team contests by simply altering a few design options, at an even lower cost. We will show the potential of these opportunities with more details in Section~\ref{sec:simulation}.

\subsection{Which Cases are Harder to Predict?}

While the the best-performing models have already improved the baseline by 24\% and generated lots of insights, the accuracy numbers do not look perfect. Indeed, individual treatment effect is perhaps one of the hardest targets for a prediction task \cite{fang2019applying}. We conduct an error analysis of the best-performing GBRT model, trying to obtain insights into what have been the harder/easier cases. 

We calculate both the prediction error ($\Delta\hat R_{C_k,j}^\text{ITE} - \Delta {R}^\text{ITE}_{C_k,j}$) and its absolute value for each driver in the test set and examine their correlations with the features, using Pearson's correlation coefficient $r$ for continuous and Student's $t$-test score for dummy features.

We find that the GBRT is less accurate when drivers have a high variance in pre-contest revenue, ($r = 0.41$, $p<.01$). This is intuitive: when the activities of a driver are irregular, their future activities are also hard to predict. This again highlights that predicting individual treatment effect is intrinsically challenging, especially in our context due to the huge heterogeneity of drivers. 
It is harder to predict for team captains than for team members ($t=12.74$, $p<.01$), and for drivers in self-formed teams than for those in system-formed teams ($t=23.07$, $p<.01$). Our model also tends to underestimate the ITEs when the average hometown distance between a driver and their teammates is larger ($r = -0.02$, $p<.01$). 

In addition, the absolute prediction error is significantly higher when there are more teams in one contest group ($t=18.93$, $p<.01$ comparing groups of 3 vs. 5 teams) and when drivers' average hourly income of the city is higher ($r = 0.23$ and $p<.01$).

\begin{figure*}[t!]
\centering
\includegraphics[width=0.85\textwidth]{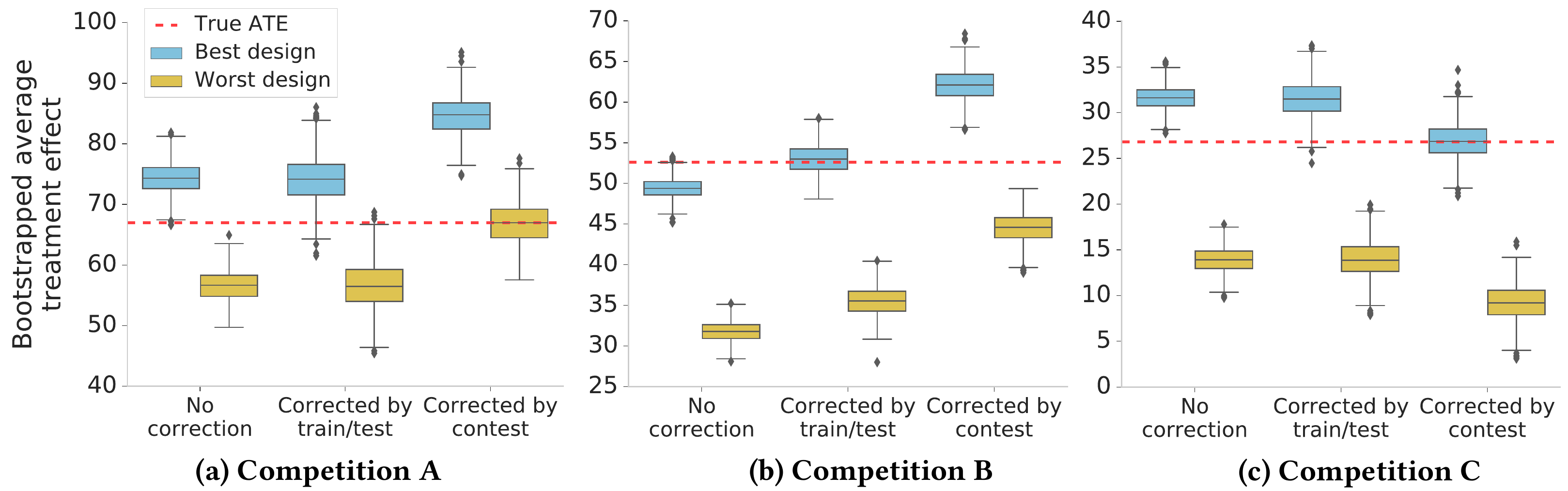}
\vskip -5pt
\caption{Simulated ATE of Three Prototype Contests under the Best Design and the Worst Design}
\vskip -5pt
\label{fig:simulation}
\end{figure*}

Overall, these factors that correlate with prediction errors are not hard to understand. Although we did not observe concerning biases, it is important to consider these patterns when applying the prediction models to different driver groups and new contexts.

%% file: sections/7-implication.tex
\section{Discussions}

\subsection{Design Implications}
We have obtained promising and actionable implications for the future design of team contests, which could affect the current practice of two aspects: \textit{contest design} and \textit{team recommender systems}.

\subsubsection{Contest Design} 
\label{sec:simulation}
Many findings about better contest design are immediately actionable. They are mostly about how to design incentives to balance the intensity and fairness of inter- and intra-team competitions. For example, (1) providing an extra bonus for the captain of the top team creates an inequality between captains and team members, which has a negative impact on the individual treatment effect; (2) excluding the lowest-performing driver from bonus allocation also results in unfair treatment within the team, which hurts the team performance in general. In a competition group, however, it is important to make sure that all teams have comparable levels of performance, so that no one loses the motivation to win. (3), it is also harmful to give awards to every team, as free lunch hinders the motivation of active competitors. These design options can all be easily reversed in future competition. To demonstrate the potential benefit of such changes, we conduct a simple counter-factual analysis through simulation.

First, we select three real contests with different choices on the three dimensions above. We hypothetically vary these design choices with everything else kept the same (such as participants, team structures, etc.), and we simulate the ``expected'' outcome through predicting the ITE of every driver in the three contests under each new design. The benefit of changing a design option can be measured by the difference between the simulated outcome under the new contest design and the outcome of the true design. Table \ref{tab:design_implication_competitions} lists the original design choices of the selected contests. 

Through simulations using the trained Lasso model, we can compare the expected outcome of the best and the worst possible configurations and the configuration in reality. Because the trained predictor is not perfect, we further adjust the prediction results by adding Gaussian noise following (1) the prediction error distribution of the training period or the test period (depending on which period the simulated contest fell into) and (2) the prediction error distribution of the original contest (with the unchanged design). Intuitively, because all other factors are controlled, we anticipate that the expected prediction error for the simulated contest shouldn't diverge much from that of the original contest. 

For each simulated competition, we bootstrap 1,000 times by sampling the number of treated drivers in the competition with replacement. Bootstrapping helps us estimate the confidence intervals of the expected average treatment effect. In Figure \ref{fig:simulation}, we report the bootstrapped average treatment effect of different simulated designs for the 3 prototype contests, including the best, the worst, and the original designs. We report the simulated ATE with prediction error uncorrected, corrected using period-level distribution, and corrected using contest-level distribution. Clearly, there is a significant difference in average treatment effect between the best and the worst design choices ($26$\%, $39$\%, and $191$\% improvement over the worst design respectively for Competition A, B, and C). In Competition A and B, the expected ATE (prediction error corrected at the contest-level) of the optimal design significantly outperforms the ATE of the original competition (using the actual design), with an improvement of as much as $26$ percent. The expected ATE does not outperform the true ATE in Competition C, as the original design is already the best. Moreover, the design choices may also affect the ROI (Revenue-over-Investment) of the competitions. As shown in Table \ref{tab:design_implication_competitions}, the ROI can increase by as much as 55\% from the original to the best design in simulation. 

The results above are promising. They demonstrate that by simply varying a few design options, both the drivers and the platform can benefit significantly. Many other design options could be improved based on the analysis in Section~\ref{sec:contest_interpret}, although it's harder to demonstrate them through a simple counter-factual simulation.

\setlength{\tabcolsep}{2pt}
\begin{table}[!ht]
\small
\vskip -5pt
\caption{\textbf{Performance of Three Prototype Contests under the Original Design and Simulated New Designs}}
\vskip -5pt
\label{tab:design_implication_competitions}
\begin{tabular}{c|cccccccc}
\hline 
 &\thead{Period} &\thead{C1} &\thead{C2} &\thead{C3} &\thead{Design} &\thead{True\\ROI} 
&\thead{Best-design ROI\\ (with 95\% C.I.)}&\thead{Worst-design ROI\\ (with 95\% C.I.)}  \\
\hline 
A &Train & Y &Y &Y &Worst & 2.86  & 4.43 (4.09, 4.76) & 2.86 (2.58, 3.13)\\
B &Test &Y  &N  &Y  &Bad & 10.61 & 13.50 (12.68, 14.30) & 10.50 (9.61, 11.34) \\
C &Train &N  &N  &N  &Best &2.58 & 2.58 (2.21, 2.94) & 0.71 (0.40, 0.99)\\
\hline
\multicolumn{9}{l}{\footnotesize{C1: Has captain bonus for top team; C2: Has team bonus for 5th team in group; }} \\
\multicolumn{9}{l}{\footnotesize{C3: Worst individual score included in team performance and bonus allocation.}}
\end{tabular}
\vskip -10pt
\end{table}
\setlength{\tabcolsep}{5pt}

\subsubsection{Team Recommendation} Our findings also shed light on how to better design team recommender systems. For example, it is better to first team up friends and former teammates, and then introduce new drivers to the team. It is beneficial to combine low-performing and newer drivers with high-performing and experienced drivers in one team. Teaming people who are from the same hometown and who work in similar areas can also boost performance.  

\subsection{Limitations}

First, this study focuses on exploring predictive factors that explain the variance of ITE across individuals, teams, contests, and cities. Although the estimation of the ITE follows the standard practice of causal inference, the prediction model does not guarantee that relations discovered between the features and the ITE are causal. Future studies are needed to establish causal relationships between the predictors identified and the ITEs. Second, we note that the benefit of optimized contest-design options is estimated based on simulations. While the three design options are carefully selected so that they are as independent as possible to other factors (so we can control the confounds), it is not impossible that changing these options may result in a change in others. For example, there is a probability that dropping the bonus for the 5th team might result in less participation. Finally, all analyses and findings are based on field experiments and data collected from one ride-sharing platform in one country. Our conclusions may be generalized to other platforms, countries, and domains with caution.

%% file: sections/8-conclusion.tex
\section{Conclusion}
We present the first predictive analysis of individual treatment effects of team competitions in DiDi, a leading platform of the ride-sharing economy. The analysis investigates hundreds of large-scale team contests in 143 cities, involving half a million drivers, tens of millions of rides, and a comprehensive set of features of the drivers, teams, contest design, and experimental conditions. Through linear and nonlinear machine learning algorithms, these features demonstrate decent predictive power of individual outcomes in team contests.  Our findings present many new insights and useful implications for the research and business practices of team competition, the sharing economy, and online field experiments in general. Some of the findings are immediately actionable in optimizing the design of upcoming team contests.  Future directions of the work include testing these insights with field experiments, investigating the causal links between the heterogeneous factors and the ITE, and generalizing the procedure to other sharing economy platforms.

%% file: sections/supplement.tex
\section{Supplement}
\subsection{Examples of Features}
Table \ref{tab:features_detailed} shows more examples of the features with implementation details. We construct more than 500 features in total, capturing contest design, driver properties, team properties, and city-level properties.

\subsection{Data Split and Model Training}
We follow the standard practice and split the contests in our analysis into training, validation, and test sets based on the time of the contests. Contests that ended on or before June 30, 2018 are used for training and contests that fell entirely in July are used as validation set.

To determine the hyperparameters, we conduct grid-search using the training and validation set. Apart from the model specific hyperparameters, we also select the best configuration of feature scales (i.e., original, Min-Max, standardization). We apply Min-Max and standardization for Lasso and Ridge, finding standardization performing the best. For GBRT models, the data of the original scale out of all three scaling methods derives the best performance.

Finally, we use all contests that ended on or before July 31, 2018 to retrain the model and report its performance on the test set, which contains the contests starting in August 2018.